\documentclass[aps,prd,amssymb,showpacs,floatfix,preprint,superscriptaddress,nofootinbib]{revtex4-1}
\usepackage{amsmath,amsfonts,graphics,epsfig,amssymb}

\begin{document}
\title{%
Constraining spacetime variations of nuclear decay rates from light curves
of type Ia supernovae }

\author{Ivan Karpikov$^{1}$, Maxim Piskunov$^{1}$, Anton Sokolov$^{2,3}$
and Sergey Troitsky}

\email{st@ms2.inr.ac.ru}

\affiliation{Institute for Nuclear
Research of the Russian Academy of Sciences, 60th October Anniversary
Prospect 7a, Moscow 117312, Russia\\
$^{2}$Physics Department, M.V.~Lomonosov Moscow State University,
Moscow 119991, Russia\\
$^{3}$Institute of Theoretical and Experimental Physics,
B.Cheremushkinskaya 25, Moscow 117218, Russia}


\date{May 12, 2015. In the original form: January 20, 2015}

\begin{center}
\begin{abstract}
The luminosity of fading type Ia supernovae is governed by radioactive
decays of $^{56}$Ni and $^{56}$Co. The decay rates are proportional to the
Fermi coupling constant $G_{\rm F}$ and, therefore, are determined by the
vacuum expectation value $v$ of the Brout--Englert--Higgs field. We use
publicly available sets of lightcurves of type Ia
supernova at various redshifts to constrain possible spacetime variations
of the $^{56}$Ni decay rate. The resulting constraint is not very tight;
however, it is the only direct bound on the variation of the decay rate
for redshifts up to $z\sim 1$. We discuss potential applications of the
result to searches for non-constancy of $G_{\rm F}$ and $v$.
\end{abstract}
\end{center}
\maketitle

\section{Introduction.}
\label{sec:intro}
Probing the limits of applicability of fundamental laws of Nature is one
of noble tasks for a present-day physicist or astrophysicist. In
particular, tests of constancy of parameters of the Standard Model of
particle physics and of General Relativity, dubbed fundamental physical
constants, attract interest for decades (see e.g.\
Refs.~\cite{const1, const2} for recent reviews and extensive lists of
references). Astrophysics and cosmology, working with largest distances
and longest time periods available for research, provide for excellent
opportunities to constrain spacetime variations of the fundamental
constants.

One of the tools to perform these studies is to compare particular
astrophysical processes taking place at various distances and in various
directions. Here, we concentrate on the latest stages of explosions of
type Ia supernovae. Being very bright, these exploding stars have been
observed at cosmologically large distances. At the same time, studies of
relatively nearby explosions revealed clear universality in features of
these supernovae and related their observable parameters to underlying
physical processes. While precise details of the explosion itself remain
to be understood, the fading luminosity of a supernova days to months
after the explosion is governed by beta decays of $^{56}$Ni and $^{56}$Co,
as it has been confirmed in 2014 by direct observations \cite{Ni, Co}.
This
fact has been used \cite{SNGF2, SNGF3} to constrain spacetime variations
of the decay rates, which are determined by the Fermi constant $G_{\rm
F}$. In the Standard Model, this ``constant'' is directly related to the
vacuum expectation value $v$ of the Brout--Englert--Higgs scalar field.
This value is determined by the effective potential for the scalar and
may, in principle, vary with the evolution of the Universe.

In this work, we take advantage of the fact that type Ia supernovae have
been extensively studied in recent years because of their importance as
distance indicators in cosmology.
We use the most recent JLA publicly available set \cite{JLA},
see \url{http://supernovae.in2p3.fr/sdss_snls_jla/ReadMe.html}.
The light curves are available for 740 objects from various
observational campaigns, spanning redshifts between 0 and $\sim 1.3$.
This enlarges significantly both the number and the redshift range of
supernovae, as compared to previous studies. This allows us to obtain the
first constraints on possible spacetime variations of the $^{56}$Ni decay
rate for the redshift range up to $z\sim 1$. They may be related to
variations in $G_{\rm F}$ and $v$ under certain assumptions.

The relation between the amount of decaying isotopes and the supernova
luminosity is not direct and we refer the reader to previous
studies \cite{SNGF2, SNGF3} for discussions of this point. The
bottom line is that after the maximum, the luminosity fading is well
described by two exponential functions, related to the $^{56}$Ni and
$^{56}$Co decays, with the slopes $\lambda$ proportional to the decay
rates $\Gamma$ with a coefficient of order one. Given considerable
universality in all type Ia supernova explosions, the coefficient of
proportionality does not vary significantly from one object to another and
constraints on the variations in the brightness slope are directly
translated into constraints on the variations of $\Gamma$.

\section{Analysis and results.}
\label{sec:analysis}
\begin{figure}
\centering
\includegraphics[width=0.75\columnwidth]{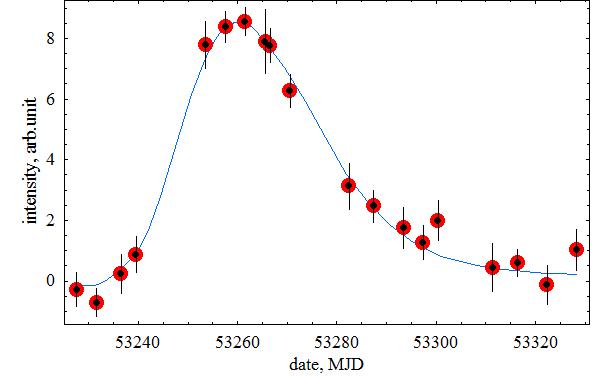}
\includegraphics[width=0.75\columnwidth]{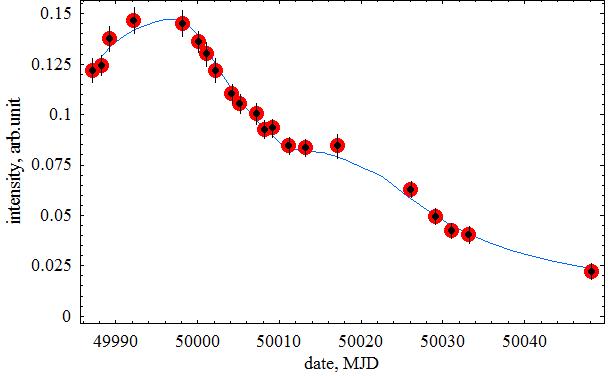}
\caption{%
Example light curves in the R band:
SNLS-04D1hy ($z=0.85$; top panel) and
SN~1995ac ($z = 0.049$; bottom panel).
The lines represent best fits obtained using
redshift-corrected templates.}
\label{fig:example-light-curve}
\end{figure}

We are dealing here with high-redshift objects, so the observations,
performed with a certain filter, correspond to various emitted
wavelengths. The relation between the luminosity and the amount of the
isotope may be wavelength-dependent, thus causing a systematic
redshift-dependent bias in our study.
For each supernova, we considered the light curve measured in the R filter
and calculated the corresponding emitted wavelength, given known $z$. The
light curve was then fitted by a standard wavelength-dependent template
obtained by interpolation of templates of Ref.~\cite{Nugent} available at
\url{https://c3.lbl.gov/nugent/nugent_templates.html}.
Example supernova light curves are
presented in Fig.~\ref{fig:example-light-curve},
together with our fit.

It is well known that type Ia supernova light curves are not precisely
universal; there exists a scatter in the time dependence of the flux
parametrized by the so-called stretch factor $s$, one of the parameters
of the fit. It is probably determined by variations in the initial chemical
composition of the exploding star. This stretch factor relates directly to
the quantity we are interested here, the decay rate, since it rescales the
time axis. Therefore, the slope of the decay exponent may be determined by
approximating a segment of the standard light curve by an exponential
function,
$F_{\rm st}(\tau) \propto \exp(-\lambda_{\rm st} \tau)$, where $F$ is
the flux and $\tau$ is the time after the maximum. We choose the B-filter
template and select the first 30 days after maximum to obtain
$\lambda_{\rm st}\simeq 0.0964$~day$^{-1}$. Then, for a supernova with a
stretch factor $s$, we have $\lambda=s \lambda_{\rm st}$ (note that the
trivial redshift multiplicative correction, $(1+z)$, is already included
at the stage of fitting).

\begin{figure}
\centering
\includegraphics[width=0.8\columnwidth]{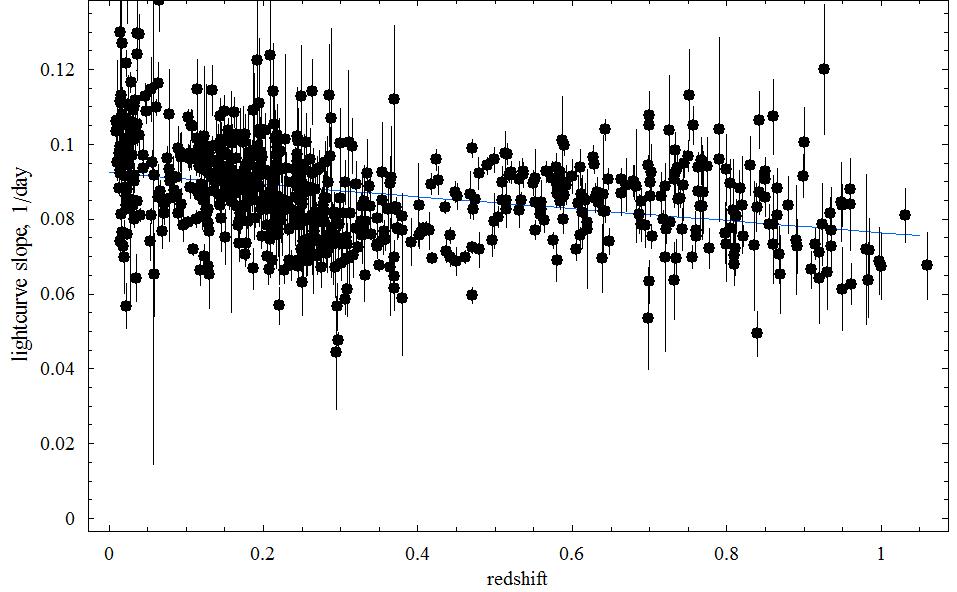}
\caption{%
The lightcurve slope $\lambda$ versus the redshift $z$ for the sample
of supernovae considered in this work.}
\label{fig:lambda-vs-z}
\end{figure}

For our analysis, we selected supernovae with at least five
data points in the light curve. We removed objects for which the fit was
poor. There remained 687
objects in the sample after applying these cuts.
Figure~\ref{fig:lambda-vs-z}
presents the redshift-corrected measured lightcurve slopes $\lambda$
versus the redshift.

In this figure, one can see the intrinsic scatter of the decay slopes,
reflecting variations in chemical composition, absorption etc. It is
manifested by the fact that the data points at a given redshift are
scattered far beyond their error bars.
We account for this intrinsic scatter by
introducing a universal systematic error, added in quadrature to the
statistical error of the slope measurement. The value of the systematic
error is adjusted in such a way that the best-fit $\chi^{2}$ of the
$\lambda(z)$ dependence
takes the median value expected for the corresponding number of degrees of
freedom\footnote{Note that for the relevant number of degrees of
freedom, there is almost no difference between median and average.}.
Assuming linear variation, $\Gamma(z)= \Gamma(0)
+\mbox{const}\times z$, we obtain
the best-fit value of the slope
in the redshift range $0 \le z \lesssim 1.0$,
$$
\frac{\Gamma(z\sim 1.0)-\Gamma(z=0)}{\Gamma(z=0)}=-0.182 \pm 0.021.
$$
We will return to the interpretation of this would-be significant effect
in Sec.~\ref{sec:concl}; here we derive a conservative constraint from it,
\begin{equation}
\left| \frac{\Gamma(z\sim 1.0)-\Gamma(z=0)}{\Gamma(z=0)}  \right|<  0.223
~~~ (95\%~{\rm CL}).
\label{Eq:*}
\end{equation}

Next, we make use of the fact that the observed objects are located in
different parts of the sky and search for global patterns of directional
variations of $\Gamma$. To this end, we study the distribution of the
lightcurve slopes $\lambda$ on the sky and fit it with a dipole harmonics
in the equatorial coordinates $(\alpha ,\delta )$,
\[
\lambda =\lambda _{0} + a_{\lambda }\, {\bf n}(\alpha ,\delta) \cdot
{\bf n}(\alpha_{0} ,\delta_{0}),
\]
where ${\bf n}(\alpha,\delta)$ is a unit vector pointing to the direction
with coordinates $(\alpha ,\delta )$; the parameters of the fit are the
monopole harmonic $\lambda _{0}$, the direction ${\bf
n}(\alpha_{0},\delta_{0})$ and the magnitude $a_{\lambda }$ of the dipole.
The dipole anisotropy is constrained as
\begin{equation}
\frac{a_{\lambda }}{\lambda _{0}}< 0.148 ~~~(95\%~CL).
\label{Eq:**}
\end{equation}

Alternatively, the first harmonic modulation in the right ascension
$\alpha $,
\[
\lambda =\lambda _{0} +A_{\lambda } \cos\left(\alpha -\alpha _{0} \right),
\]
is constrained as
\begin{equation}
\frac{A_{\lambda }}{\lambda _{0}}< 0.051 ~~~(95\%~CL).
\label{Eq:***}
\end{equation}
The constraints (\ref{Eq:*}), (\ref{Eq:**}), (\ref{Eq:***}) are the main
direct results of our study.

\section{Discussion and conclusions.}
\label{sec:concl}
The constraints on the variation of the $^{56}$Ni decay rate obtained
here may be compared to other (scarce) limits on the variation of
beta decay rates, see Table~\ref{tab:constraintsR}.
\begin{table}
\begin{center}
\begin{tabular}{ccccc}
\hline
\hline
Isotope & Method & Result & Redshift & Reference\\
\hline
$^{56}$Ni & supernovae & $>-0.223$ & 1.0 & this work\\
$^{56}$Co & supernovae & $-0.272 \pm 0.156$ & 0.024 &  \cite{SNGF3}
\\
$^{187}$Re& meteorites & $-0.016 \pm 0.016$ & 0.45 & \cite{meteorite}
 \\
\hline
\hline
\end{tabular}
\end{center}
\caption{Comparison of astrophysical constraints on variations of
nuclear beta decay rates.}
\label{tab:constraintsR}
\end{table}
The next step, that is to relate the observed result to the constancy of
fundamental constants, is not straightforward and requires additional
assumptions (see e.g.\ an excellent description of how contrived this step
is in Ref.~\cite{Planck}, Sec.~2.2, not to mention numerous more general
discussions in the literature).

First, the variations may be constrained in a self-consistent way for
dimensionless constants only, while the decay rates and related, more
fundamental, parameters $G_{\rm F}$ and $v$ are all dimensionful.
A plausible dimensionless combination for our analysis may
be formed with the help of the gravitational constant $G$, cf.\ the
``$\mu_{\rm H}$'' constant of Ref.~\cite{Duff2014}. This raises the second
subtle point, which
constants are allowed to vary. For instance,
allowing for variations of the gravitational constant may change the
conclusions significantly. Therefore, constraining just the variation of a
constant is not that useful; a particular physical model predicting the
variations of the entire set of fundamental constants should be
constrained instead. The results presented here, that is constraints on
the variations of a physical quantity, the decay rate, may be used in any
work of that kind.

However, just to put our results in context and to
compare them with other limits, we interpret our bounds on the variation
of $\Gamma$ in terms of $v$ within a simplified assumption that no other
constants are allowed to vary. This is presented for demonstration only,
because we do not mean any particular underlying theory behind the
assumption.

Apart from the approach we use here,
there exist other ways to constrain variations
of $v$. The change of another radioactive decay rate, also proportional to
$G_{\rm F}$, was constrained with meteorite data. People analyzed the
effect of $v$ on the Big Bang nucleosynthesis (BBN, through several
quantities) and on the Cosmic Microwave Background (CMB) and the
Large-Scale Structure (LSS) of the Universe (through the change in the
electron mass which is proportional to $v$). All indirect constraints
require model-dependent interpretation which is especially hard when
effects of the strong interaction, which are not calculable in the
fundamental theory, are taken into account. For completeness, we mention
here also the possibility to constrain variations of $v$ through the
change in the electron mass $m_{e}$ by constraining variations of the
electron-proton mass ratio $\mu$, the most restrictive constraints
coming from measurements with atomic clocks (AC) and with quasar (QSO)
spectra. These result may constrain variations of $v$ if it is assumed that
the proton mass does not change with the change of $v$ while the electorn
mass does, because the former is dominated by strong-coupling effects and
is not given by the sum of quark masses, which are in turn proportional to
$v$ in the high-energy limit. This point requires, however, a quantitative
derivation which is presently missing. A far from being exhaustive list of
recent constraints and references is presented in
Table~\ref{tab:constraints}
\begin{table}
\begin{center}
\begin{tabular}{ccccc}
\hline
\hline
Quantity & Method & Result & Redshift & Reference\\
\hline
$\Delta v/v$ & supernovae & $<0.056$ & 1.0 & this work\\
             & supernovae & $0.068 \pm 0.039$ & 0.024 & \cite{SNGF3}
 \\
             & meteorites & $0.004 \pm 0.004$ & 0.45 & \cite{meteorite}
 \\
             & BBN        & $0.0134 \pm 0.0014$ &
$\sim10^{8}$&\cite{bbn2}
 \\
& BBN        & $0.004 \pm 0.002$ &  $\sim 10^{8}$&\cite{bbn1}
 \\
\hline
$\Delta m_{e}/m_{e}$ & CMB &$0.004 \pm 0.011$ &  $\sim
10^{3}$&\cite{Planck}
 \\
                     & LSS &$0.006^{+0.014}_{-0.016}$ &  $\sim
10^{3}$&\cite{lss}
 \\
\hline
$\Delta\mu/\mu$ & QSO & $(0.0 \pm 1.0)\times 10^{-7}$ & 0.89&
\cite{qsoR1}
 \\
                & QSO & $(-2.4\pm 2.3)\times 10^{-7}$ & 0.89&
\cite{qsoR4}
 \\
                & QSO & $(0.0\pm 1.5)\times 10^{-6}$ & 1.3& \cite{qsoR3}
 \\
                & QSO & $(-7.6\pm 8.1)\times 10^{-6}$ & 2.4&
\cite{qsoR5}
 \\
                & QSO & $(-1.7\pm 1.7)\times 10^{-6}$ & 3.17&
\cite{qsoR2}
 \\
\hline
$\dot{\mu}/\mu$ &AC& $(-0.5\pm 1.6)\times
10^{-16}$~yr$^{-1}$&0&\cite{atom2}
 \\
                &AC&$(0.2\pm 1.1)\times
10^{-16}$~yr$^{-1}$&0&\cite{atom1}
 \\
\hline
\hline
\end{tabular}
\end{center}
\caption{Comparison of recent constraints on variations of $v$,
$m_{e}$ and $\mu$ obtained by various methods; see text for details.}
\label{tab:constraints}
\end{table}
%

Studies of the light-curve shapes of type Ia supernovae, in particular of
their redshift dependence, have a long story. Probably the first work on
this subject was one by Pskovskii~\cite{Pskovskii}. Studies of the rise
time of type Ia supernovae, whose redshift dependence was discussed, in
particular, in Ref.~\cite{2006AJ132-1707}, can hardly be directly used for
the purposes of the present work. Modern studies of the fading part of the
light curve, e.g.\ \cite{2007ApJ667L37}, suggest a trend to decreasing of
the decay slope of the curve as the redshift increases, similarly to what
we observe in the present work. A possible explanation is that the slope
depends on the environment in the supernova host galaxy: the decay of the
light curve is slower in young stellar environments, while the age of
stellar populations decreases with increasing redshift. Therefore,
selecting only supernovae residing in galaxies with active star formation
may result in a more homogeneous sample \cite{1409.2951} and thus reduce
the corresponding bias in the study of the decay rates. This is a
prospective way to improve the precision of constraints presented here.

To conclude, we have obtained constraints on the spacetime variation
of the $^{56}$Ni decay rate; our results are given in Eqs.\ (\ref{Eq:*}),
(\ref{Eq:**}), (\ref{Eq:***}). These are the first bounds on the variation
of the nuclear decay rates obtained for the redshift range as large as
$0\le z \lesssim 1$, see Table~\ref{tab:constraintsR} for comparison
with other results. The interpretation of the results in terms of varying
fundamental constants like the Higgs scalar vacuum expectation value $v$
may require a formulation of a particular theory describing how these, and
other, constants vary.
Within simplified assumptions, we
compared our result in terms of the variation of $v$ to others, including
more model-dependent ones, in Table~\ref{tab:constraints}.

\begin{acknowledgments}
We are indebted to the anonymous referee for helpful comments
and to M.~Pshirkov for interesting discussions. The work has made use of
JLA supernova lightcurves publicly available at
\url{http://supernovae.in2p3.fr/sdss_snls_jla/ReadMe.html}. The work of
M.P.\ and S.T.\ was supported in part by the Russian Foundation for Basic
Research, grant 13-02-01293.
\end{acknowledgments}

\end{document}